\documentclass[twocolumn]{article}

\usepackage{fullpage, times, mathptm, graphicx, natbib}

\begin{document}

\title{Scaling in the growth of geographically subdivided populations:
  invariant patterns from a continent-wide biological survey\thanks{In
    press, \textit{Philosophical Transactions of the Royal Society of
      London, Series B}.}}
 
\author{Timothy H.~Keitt,$^1$ Luis A.~N.~Amaral,$^2$ Sergey
  V.~Buldyrev$^2$ \\ and H.~Eugene Stanley$^2$ \bigskip \\ $^1$
  Department of Ecology and Evolution \\ State University of New York
  at Stony Brook, Stony Brook, NY 11794 USA \medskip \\ $^2$ Center
  for Polymer Studies and Department of Physics \\ Boston University,
  MA 02215 USA}

\maketitle

\begin{abstract} 
  
  We consider statistical patterns of variation in growth rates for
  over 400 species of breeding birds across North America surveyed
  from 1966 to 1998.  We report two results.  First, the standard
  deviation of population growth rates decays as a power-law function
  of total population size with an exponent $\beta = 0.36\pm0.02$.
  Second, the number of subpopulations, measured as the number of
  survey locations with non-zero counts, scales to the 3/4-power of
  total number of birds counted in a given species.  We show how these
  patterns may be related, and discuss a simple stochastic growth
  model for a geographically subdivided population that formalizes the
  relationship.  We also examine reasons that may explain why some
  species deviate from these scaling-laws.

\end{abstract} 

\section{Introduction}

Perhaps one of the most intriguing patterns in ecology is Taylor's
law~\citep{anderson-etal-1982, soberon-loevinsohn-1987,
  routledge-swartz-1991, leps-1993, curnutt-etal-1996, maurer-1999}.
Taylor (\citeyear{taylor:lr-1961}) was the first to notice that when
the mean $\langle S \rangle$ of a population survey is plotted versus
its variance $\sigma^2(S)$, either in space or time, the
relationship is typically a power-law with a fractional exponent
\begin{equation}
\sigma^2(S) \sim \langle S \rangle^{\gamma}\,.
\label{eq:taylor}
\end{equation}
Taylor was originally interested in the slope of the power-law
relationship as a scale-free measure of spatial contagion or
dispersion --- values greater (less) than one indicate spatial
clustering (over-dispersion).  Later, Taylor used both spatial and
temporal scaling as a basis for comparative studies of, in his words,
``synoptic population dynamics'' across taxonomic
groups~\citep{taylor:lr-woiwod-1982, taylor:lr-1984}.

Taylor's synoptic approach is, in many respects, a precursor to the
recent development of ``macroecology,''~\citep{brown-1995}, a
sub-discipline of ecology and
biogeography~\citep{macarthur-wilson-1967} that seeks to identify
broad patterns in species' abundance and distribution.  Macroecology
has largely focussed on static patterns, such as spatial relationships
between abundance and environmental factors~\citep{brown-etal-1995}
and relationships between metabolic energy use and geographic
distribution~\citep{brown-maurer-1987}.  Thus, relatively few
continental-scale macroecological studies~\citep{maurer-1994,
  maurer-1999} have explored interactions between spatial distribution
and population variability through time.

In this paper, we adopt Taylor's synoptic approach and analyze one of
the most comprehensive macroecological data sets available, the North
American Breeding Bird Survey~\citep{peterjohn-1994}.  The data are
estimates of local abundance (counts) for over 600 bird species
recorded at 2--3 thousand sites (routes) across North America for the
years 1966 to 1998.  Unlike many previous macroecological studies, our
focus is on linking geographic distribution to population dynamics.
We report two new scaling laws, closely related to Taylor's power-law,
for these data: one relating variability of population time series to
their mean, and another relating number of sites occupied to total
population size.  In addition, we show how these patterns may be
related, and discuss a simple stochastic growth model for a
geographically subdivided population that formalizes the relationship.

\section{Scaling of species growth rates}

\begin{figure*}[t]
  \begin{center}
    \begin{tabular}{cc}
      \includegraphics[height=2.5in]{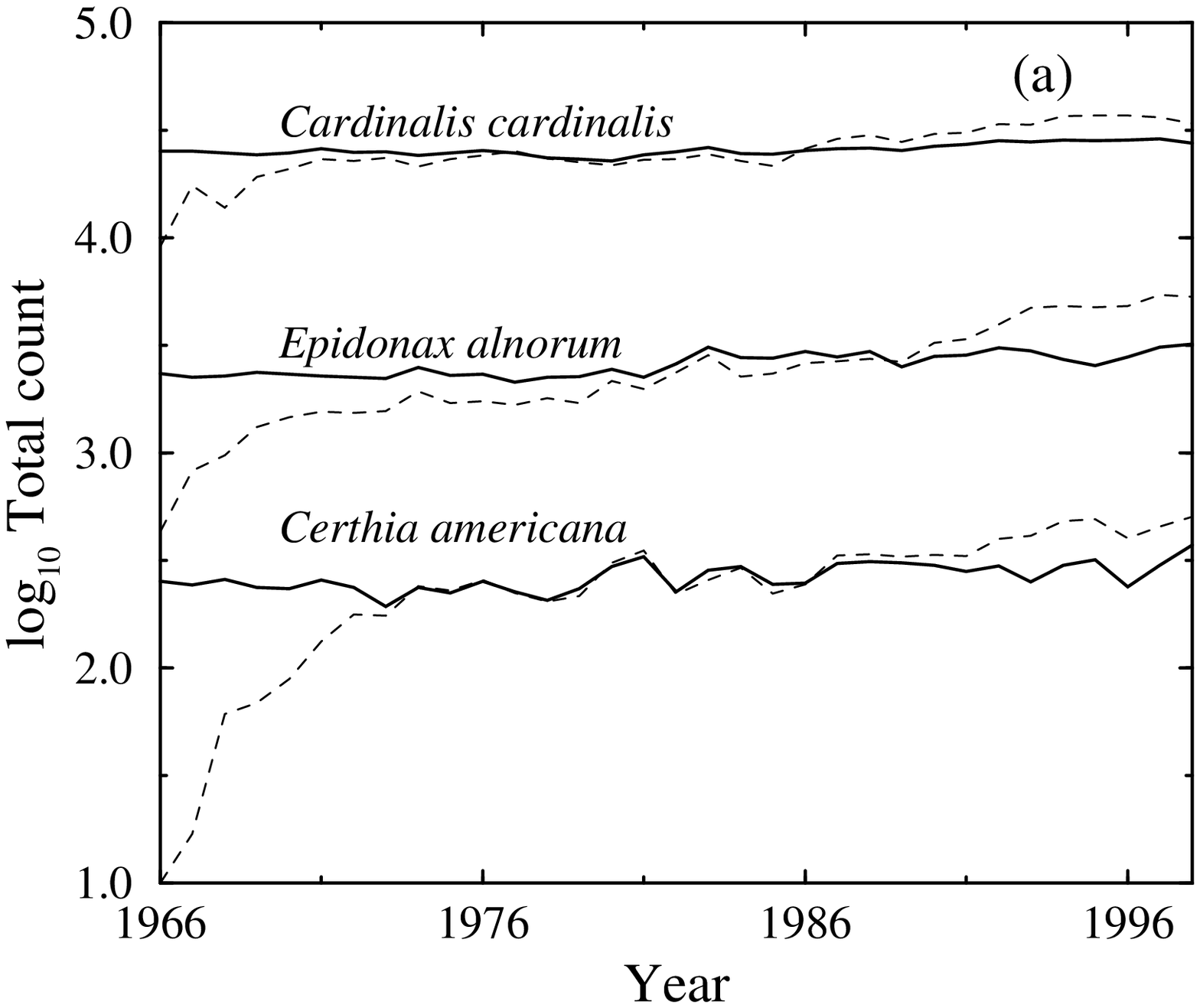} &
      \includegraphics[height=2.5in]{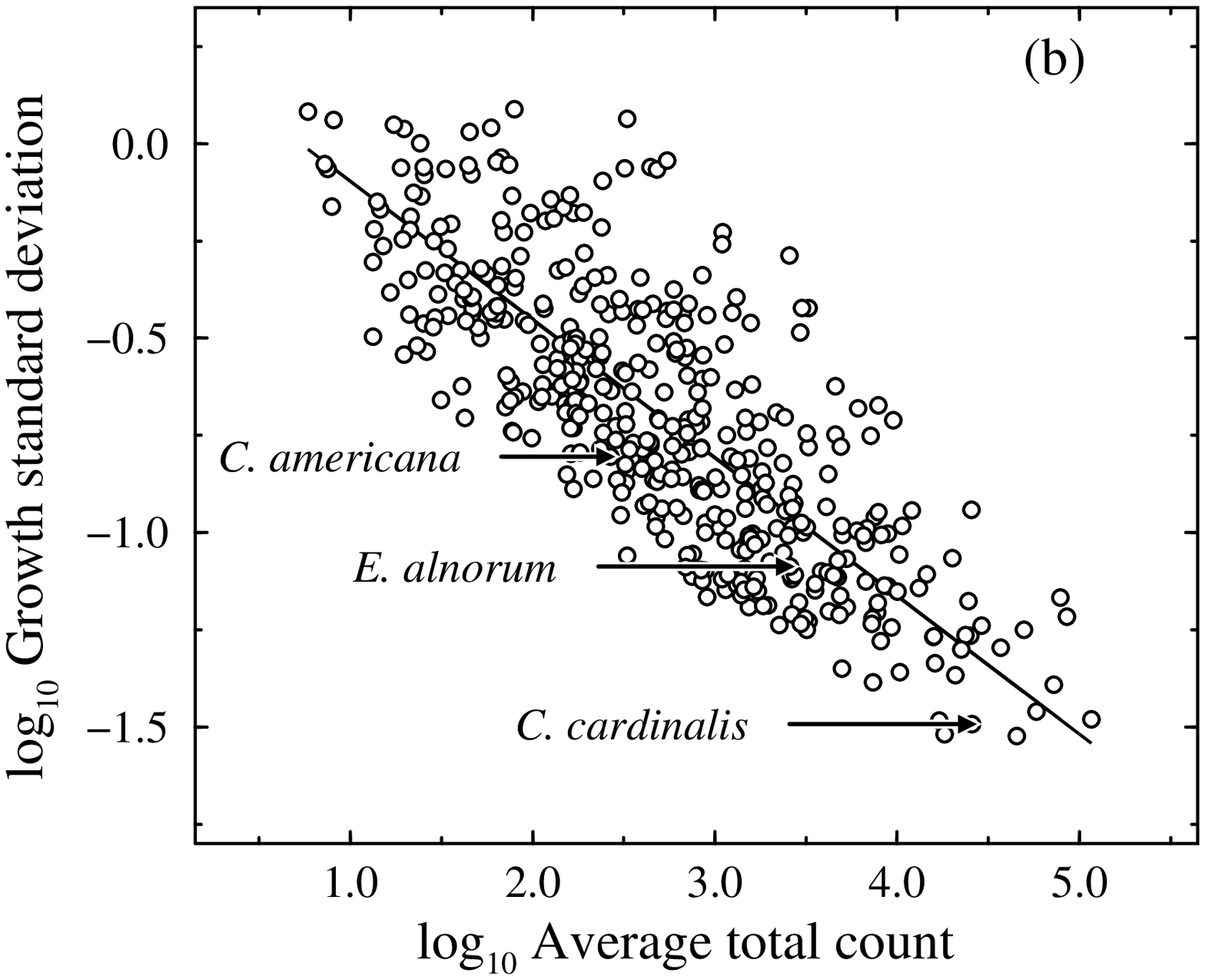}
    \end{tabular}
    \caption{Scaling of ecological time
      series.  \textbf{(a)} Time series of total counts for three
      species of birds (Northern Cardinal, Alder Flycatcher and Brown
      Creeper).  Dashed lines show the uncorrected total number of
      birds counted in a given year.  Solid lines are bias-corrected
      totals (see text for explanation).  \textbf{(b)} Standard
      deviations of empirically observed population growth rates
      plotted against average annual total counts for 428 species of
      North American breeding birds.  Results for the time series
      plotted in panel \textbf{(a)} are highlighted.  Notice that the
      highlighted time series are less variable than other species in
      the data set with similar total population
      size.\label{fig:time_series_scaling}}
  \end{center}
\end{figure*}

Our goal is to understand temporal variation in abundance at the
entire population level.  We therefore compute time series of total
counts for each species by summing over all routes surveyed in a given
year (Fig,~\ref{fig:time_series_scaling}a).  (We have previously
analyzed these data at the individual route level, see Keitt and
Stanley, \citeyear{keitt-stanley-1998}.)  The aggregated time series
should be relatively robust to observer errors inherent in the
route-level data~\citep{kendall-etal-1996} since random under-counts
and over-counts will cancel in the summation.  After performing a bias
correction (see Appendix for details), the resulting time series are
relatively free of systematic trends, particularly in the early years
of the survey when the number of routes was increasing rapidly
(compare dashed to solid lines in
Fig.~\ref{fig:time_series_scaling}a).

We choose as our measure of the magnitude of time-series fluctuations,
the logarithm of the ratio of successive counts,
\begin{equation}
g(t) \equiv \log \left( \frac{S(t+1)}{S(t)} \right) \,,
\label{e.grate}
\end{equation}
where $S(t)$ and $S(t+1)$ are the total number of birds of a given
species counted in years $t$ and $t+1$, respectively.  This measure
has several nice properties.  First, any multiplicative,
time-independent sample bias cancels in the ratio.  Second, the
measure has a natural interpretation in terms of population demography
since, in a closed population, $\exp[g(t)] \approx 1$ + (per capita
birth rate - per capita death rate).

As shown in Fig.~\ref{fig:time_series_scaling}b, the standard
deviation $\sigma(g)$ of population growth rates is strongly related
to the average total population size.  The relationship follows a
power-law
\begin{equation}
\sigma(g) \sim \langle S \rangle^{-\beta} \,
\label{e.sigma}
\end{equation}
for over four orders of magnitude in $\langle{S}\rangle$, the total
count averaged across years.  For these analyses, we are not
interested in predicting a ``dependent'' variable from an
``independent'' variable.  Rather, we are interested in modeling the
functional form of the interdependence among variables.  We therefore
use major-axis regression~\citep{sokal-rohlf-1995} to estimate model
parameters.  Major-axis regression is based on computing the leading
eigenvector of the covariance matrix, and minimizes squared errors
measured perpendicular to the trend line.  We also restrict our
analysis to non-zero time series of at least 25 years in length and
with a minimum average total count of no fewer than 5 birds per year.
Using major-axis regression with bootstrap precision estimates, we
find $\beta = 0.36\pm0.02$.  Taylor's exponent (replicated across
species) is simply $\gamma = 2(1-\beta) = 1.44\pm0.04$.

\begin{figure*}[t]
  \begin{center}
    \begin{tabular}{cc}
      \includegraphics[height=2.5in]{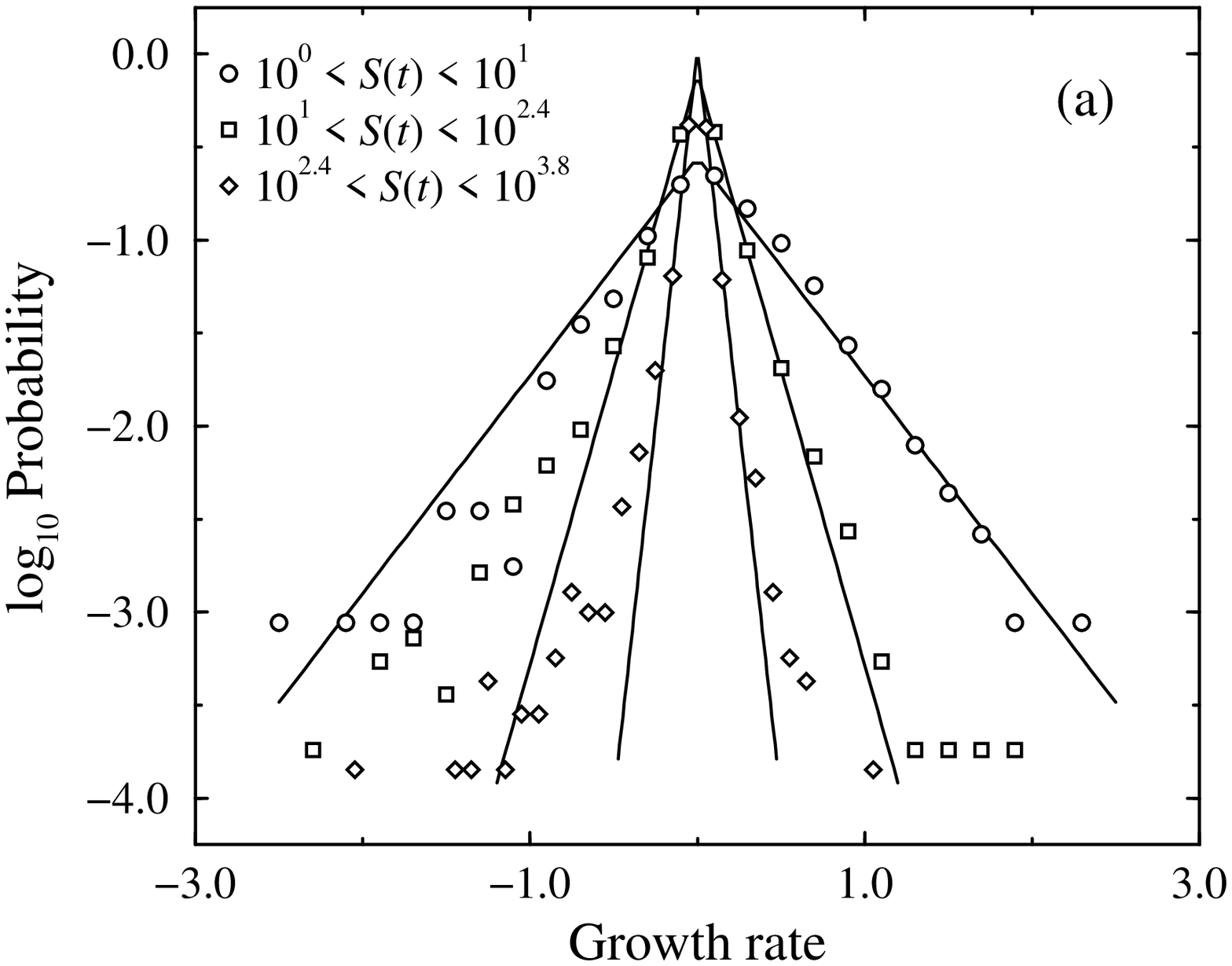} &
      \includegraphics[height=2.5in]{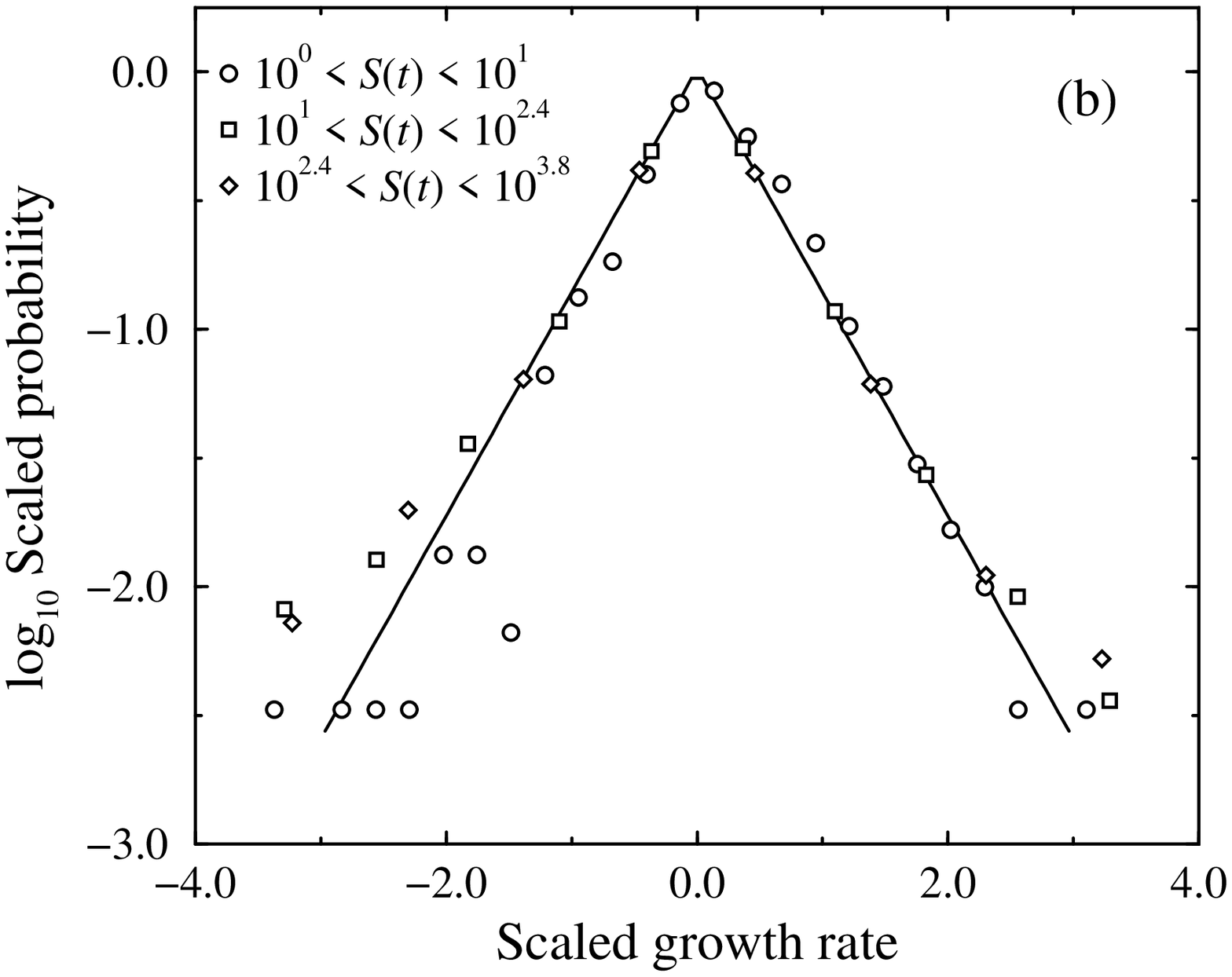}
    \end{tabular}
    \caption{Distribution of growth rates for
      different population size classes.  \textbf{(a)} Probability
      density $p(g|S)$ of the growth rate $g$ for all bird species in
      the North American Breeding Bird Survey database. The
      distribution represents all annual growth rates observed in the
      31-year period 1966--1996.  We show the data for three different
      bins of initial sizes.  The solid lines are exponential fits to
      the empirical data close to the peak.  The approximately
      triangular shape of the distribution may be the result of mixing
      Gaussians with different widths~\citep{amaral-etal-1998}.
      \textbf{(b)} Scaled probability density $p_{\mbox{\scriptsize
          scal}} \equiv \sigma p(g|S)$ as a function of the scaled
      growth rate $g_{\mbox{\scriptsize scal}} \equiv [g-\bar g] /
      \sigma$ for all species and years in the survey.  The values
      were rescaled using the measured values of $\bar g$ and
      $\sigma$.  All the data collapse upon the universal curve
      $p_{\mbox{\scriptsize scal}} = f (-|g_{\mbox{\scriptsize
      scal}}|)$. \label{fig:tent_shape}} 
  \end{center}
\end{figure*}

We also study the scaling properties of population growth rates by
examining changes in the distribution of growth rates with increasing
population size, a technique familiar to statistical physicists.
First, we separate the observed growth rates into three bins according
to the total count $S(t)$ and then construct a histogram to estimate
the conditional probabilities $p(g|S)$.  The resulting distributions
are roughly triangular in shape with the width depending on $S$
(Fig.~\ref{fig:time_series_scaling}a).  (The triangular shape may
result from summing over a large number of time series with different
local variances, see Amaral et al.,~\citeyear{amaral-etal-1998}).  If
the distributions are ``self-similar'' (i.e., exhibit scaling), then
we should be able to identify a function $f$ that rescales the
distributions so that they ``collapse'' onto each
other~\citep{stanley:he-1971}.  We plot the scaled quantities:
\begin{equation}
\sigma(S) ~ p \left( \left . \frac{g}{\sigma(S)} \right| S \right) ~~~{vs.}
~~~ \frac{g}{\sigma(S)} \,,
\label{e.scaled}
\end{equation}
(Fig.~\ref{fig:tent_shape}a) and find that indeed the three curves do
collapse onto each other (Fig.~\ref{fig:tent_shape}b), suggesting that
$p(g | S)$ follows a universal scaling form
\begin{equation}
p(g | S) \sim \frac{1}{\sigma(S)} ~~ f \left( \frac{g}{\sigma(S)}
\right)\,.
\label{eq:dist-grate}
\end{equation}

These results are interesting for a number of reasons.  In statistical
physics, the presence of non-trivial scaling is usually taken to mean
that the dynamics are largely governed by simple geometric properties
of the system and do not depend strongly on detailed properties of the
system subcomponents~\citep{wilson:kg-1983}.  Thus, it is remarkable
that we should find strong evidence for scaling across such a
taxonomically and ecologically diverse set of species as found in the
North American Breeding Bird Survey.  These results suggest that the
dynamics of North American Breeding Birds are unexpectedly ``simple''
and depend primarily on common patterns of internal population
structure across species ranges, rather than details of individual
species life-histories.

\section{Spatial structure of subpopulations}

Another reason the growth-scaling results are interesting is the large
variability of highly abundant species.  Imagine the null model that
each population is subdivided into $n$ equally sized, independent
subpopulations, and that the number of these subpopulations depends
linearly on $S$.  The expectation, according to the central limit
theorem, is that the standard deviation in growth rates should decay
as the -1/2 power of $S$.  The observed decrease in fluctuations is
considerably slower (i.e., $\beta < 1/2$), such that highly abundant
species are considerably more variable than expected under the null
model.

To account for the increased variability for highly abundant species,
we require that the number of subpopulations does not scale in a
simple, linear fashion with increasing $S$, but instead takes the form:
\begin{equation}
n \sim S^{1-\alpha}\,,
\label{eq:number-of-units}
\end{equation}
with $\alpha \ne 0$.  Values of $\alpha > 0$ will be found, for
example, when the ``typical'' size of the subpopulations also scales
with total abundance, i.e., there is a positive relationship between
regional and local abundance, a well documented pattern in
macroecology~\citep{gaston-lawton-1988, gaston-1996}.  The positive
correlation between local and regional abundance results in fewer
subpopulations for a given total population size as each subpopulation
accounts for more individuals.  Again appealing to our observation
that, under the central limit theorem, $\sigma(g) \sim n^{-1/2}$, and
in combination with equations (\ref{e.sigma}) and
(\ref{eq:number-of-units}), it is strait forward to show that for
roughly equal-sized subpopulations, the estimated exponents must obey
\begin{equation}
\beta = \frac{1 - \alpha}{2} \,.
\label{eq:beta-independent}
\end{equation}

\begin{figure*}[t]
  \begin{center}
    \begin{tabular}{cc}
      \includegraphics[height=2.5in]{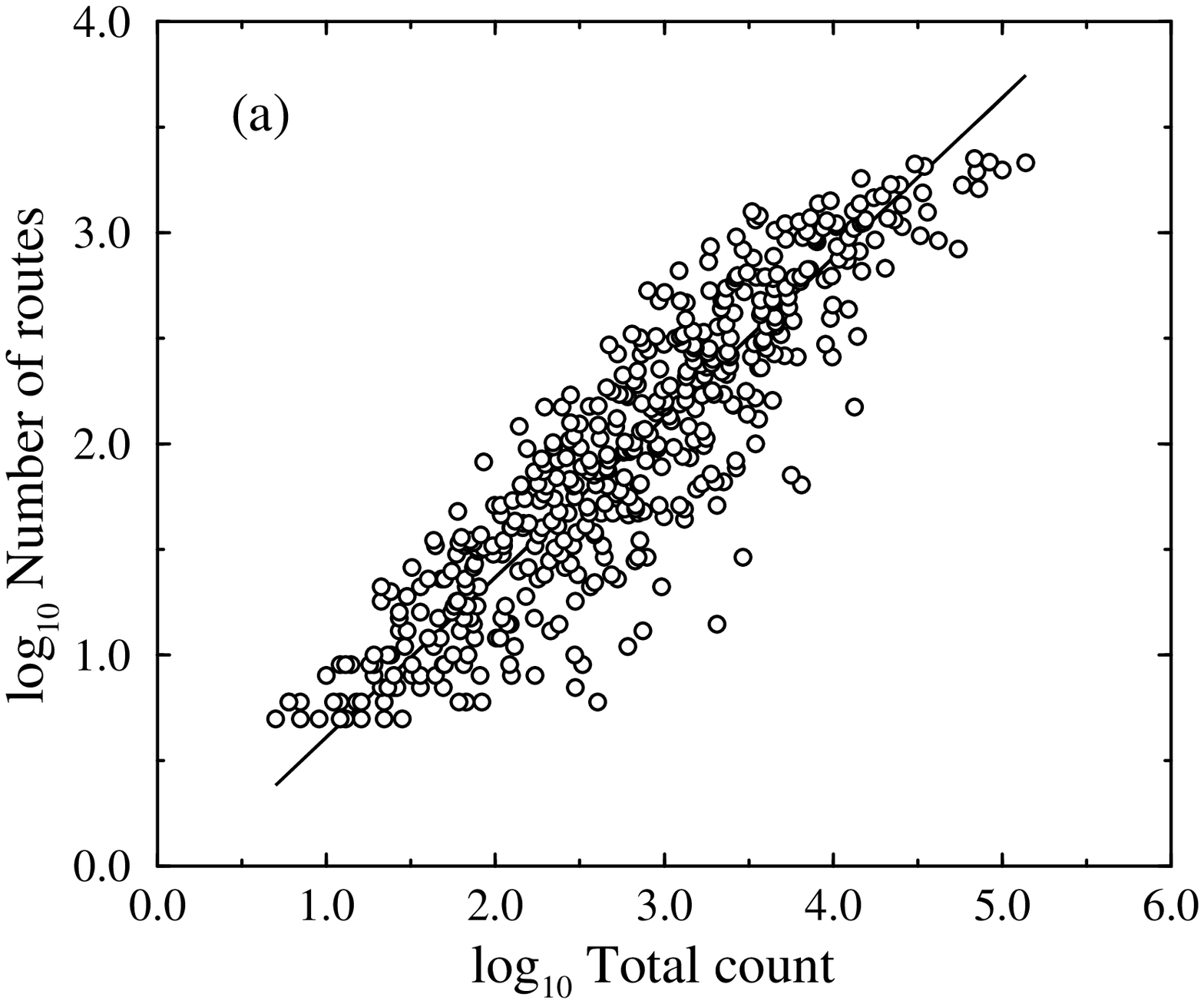} &
      \includegraphics[height=2.5in]{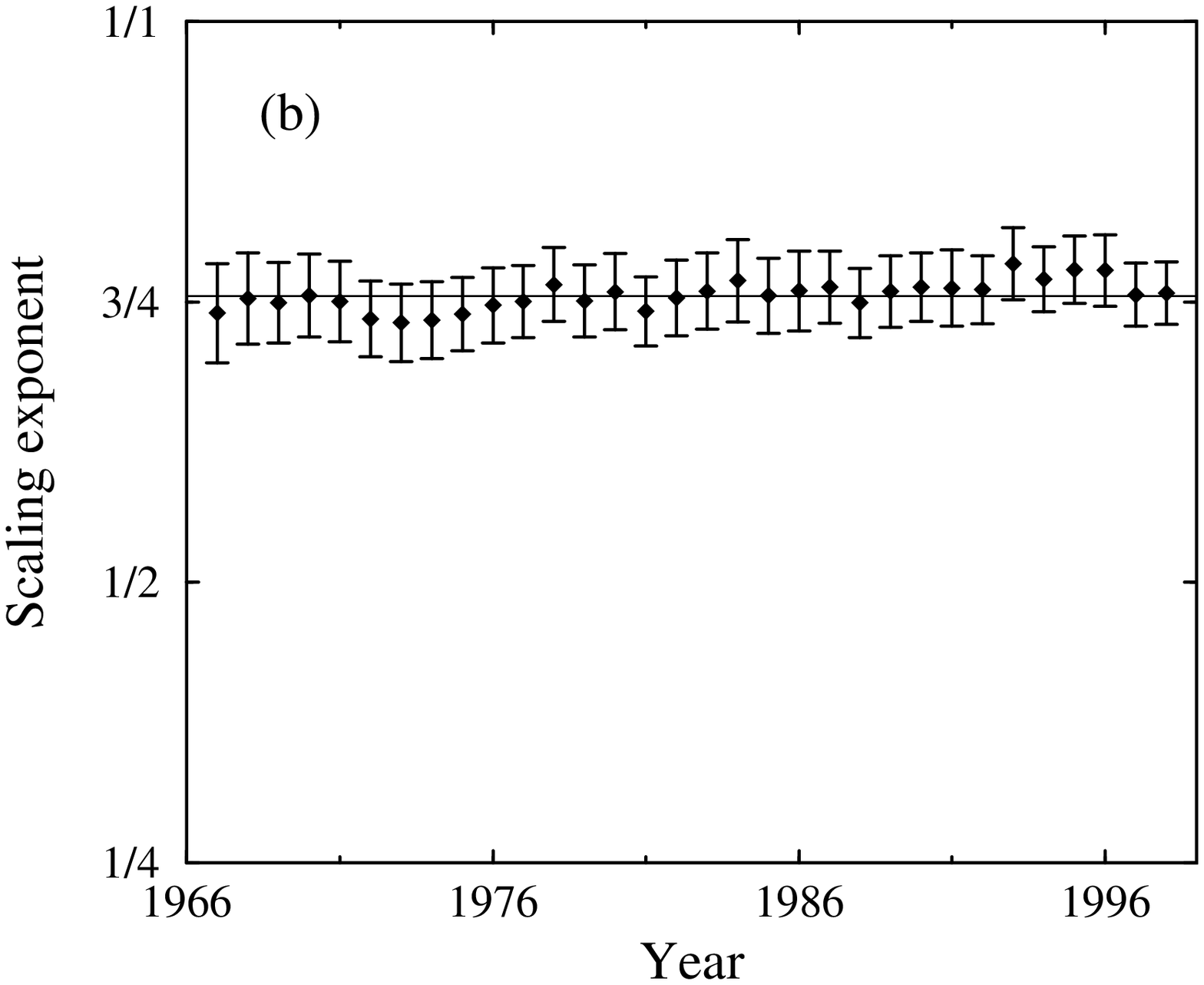}
    \end{tabular}
    \caption{Statistical analysis of the number
      of routes populated by a given bird species. \textbf{(a)} Double
      logarithmic plot of the number of routes with non-zero counts
      $\tilde{n}(t)$ versus total number of birds counted $S(t)$ for
      each species observed in 1997.  The bias correction applied to
      the time series in Fig.~\ref{fig:time_series_scaling} is
      unnecessary in this case as all data come from a single year.
      The data for all species follow closely a straight line the
      log-log plot suggesting a power law dependence.  From the slope
      of the line, we estimate $1-\alpha = 0.75 \pm 0.03$.
      \textbf{(b)} We perform a similar analysis for all 31 years in
      the database and plot the exponent estimates for each of the
      years.  Our results show that the power law dependence remains
      remarkably stable during the 31 survey year, clustering around
      $\alpha = 0.25$.  Error bars are bootstrap 95\% confidence
      intervals.\label{fig:bbs_units_scaling}}
  \end{center}
\end{figure*}

We do not have access to a precise estimate of the number of
subpopulations for each species in the survey. However, we can use as
a proxy the number of survey routes where a species had a non-zero
count in a given year.  To test the assertion in
Eq.~\ref{eq:beta-independent}, we plot number of survey routes with
non-zero counts $\tilde{n}(t)$ versus the (uncorrected) total count
$S(t)$ for all bird species recorded in the survey in 1997, excluding
species seen at fewer than 5 routes or with fewer than 5 total
individuals counted.  The data follow closely the power law dependence
predicted by Eq.~(\ref{eq:number-of-units}) with an exponent $\alpha =
0.25 \pm 0.03$, again using major-axis regression with bootstrap
precision estimates (Fig.~\ref{fig:bbs_units_scaling}a).  Remarkably,
the estimate of $\alpha$ predicts a value of $\beta = 0.38 \pm 0.02$,
very close to the estimate ($\beta = 0.36 \pm 0.02$) obtained by
measuring the standard deviation in growth rates directly
(Fig.~\ref{fig:time_series_scaling}b).  Even more striking is the
consistency of our estimate of $\alpha$ across years, despite large
changes in the number and spatial distribution of sampling locations
through time (Fig.~\ref{fig:bbs_units_scaling}b).  These results
directly imply that average local abundance $\langle s \rangle =
S(t)/\tilde{n}(t)$, scales with total (regional) abundance according
to $\langle s \rangle \sim S(t)^\alpha$.

\begin{figure*}[t]
  \begin{center}
    \begin{tabular}{cc}
      \includegraphics[height=2.5in]{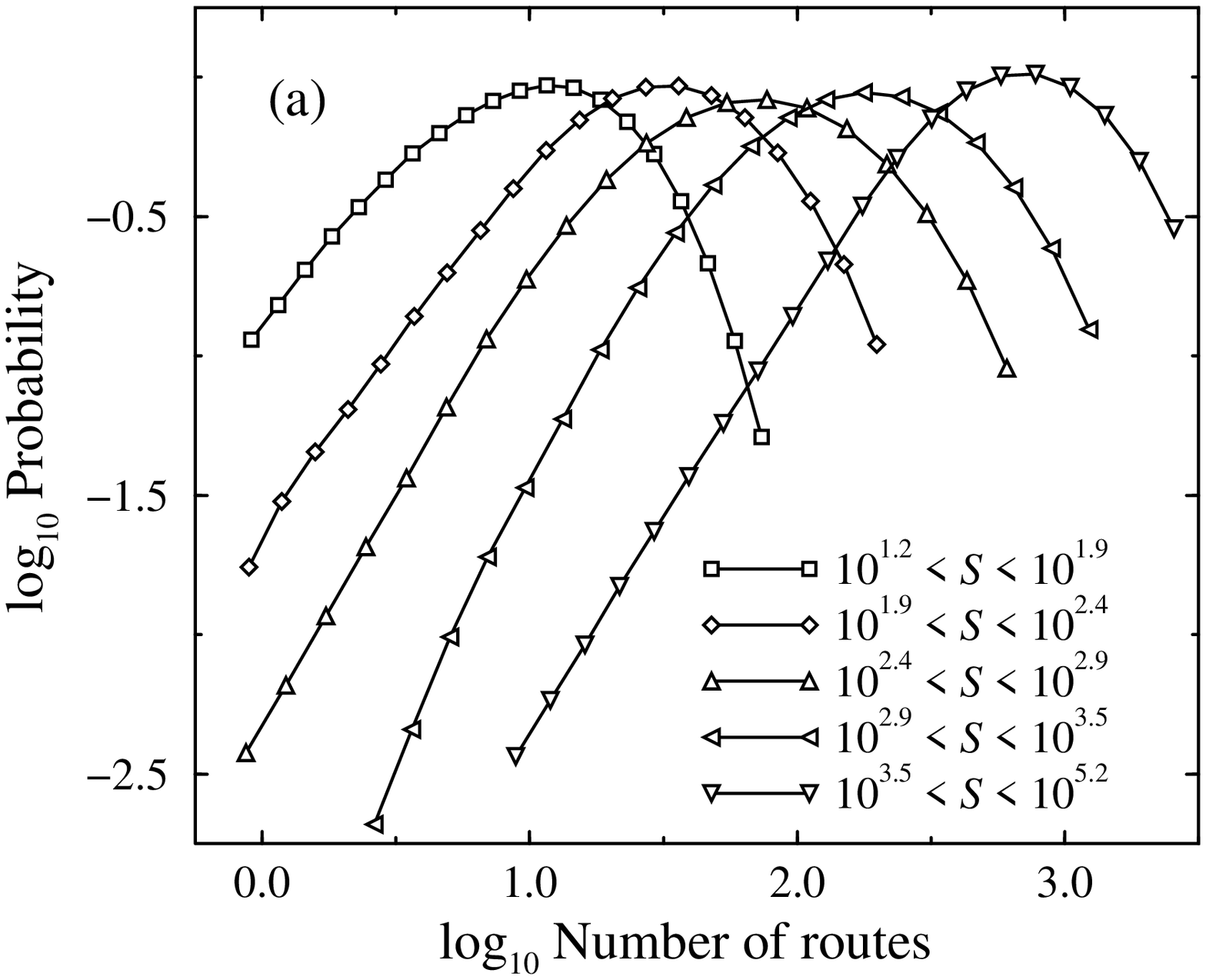} &
      \includegraphics[height=2.5in]{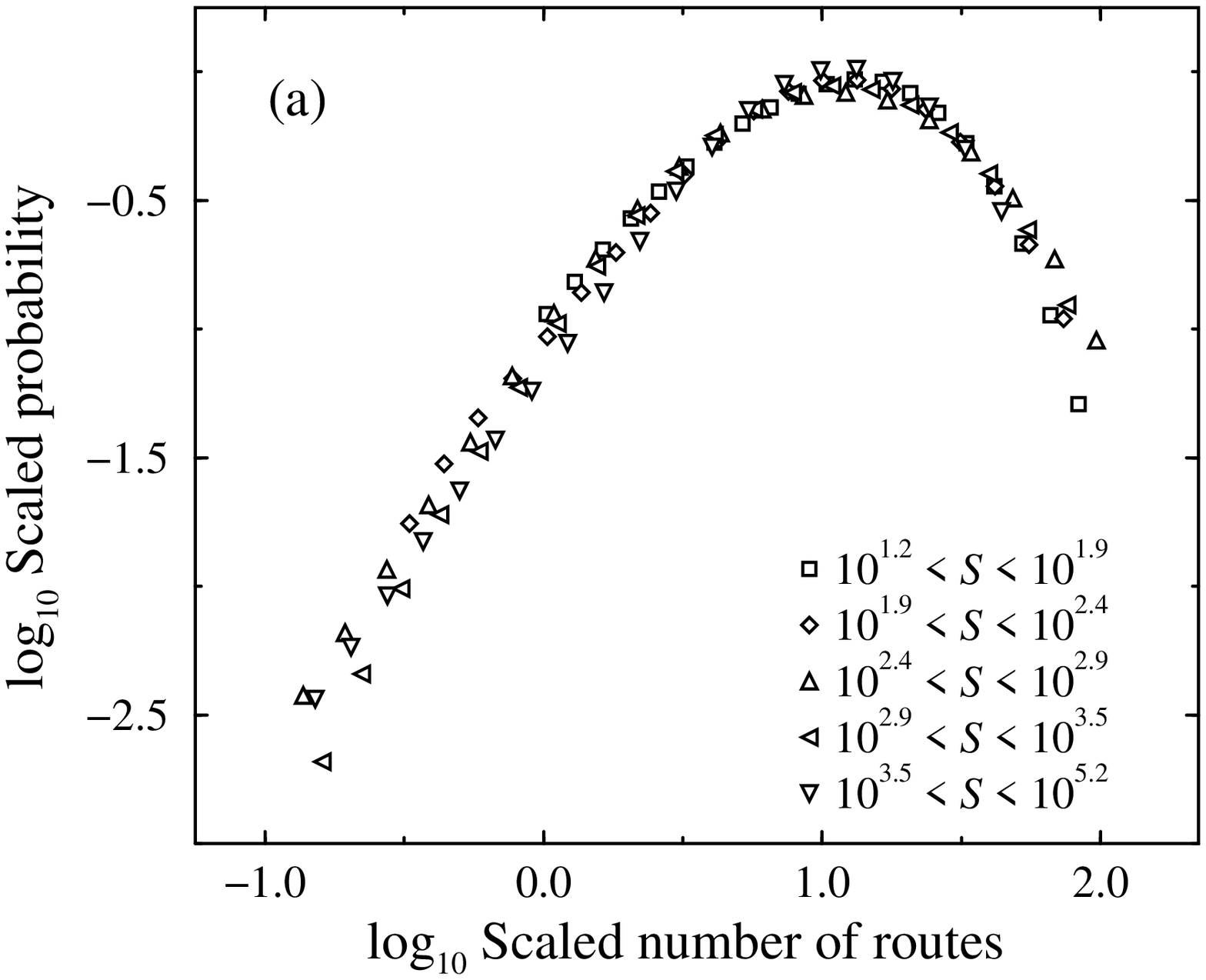}
    \end{tabular}
    \caption{Statistical analysis of the number of
      routes populated by a given bird species. \textbf{(a)}
      Conditional probability density function $\rho(\tilde{n}|S)$ of
      finding $\tilde{n}$ non-zero count routes for a bird species
      with $S$ total counts. To improve the statistics, we partition
      the bird species into four groups according to size.
      \textbf{(b)} To illustrate the scaling relation
      (\protect\ref{eq:dist-scaling}), we plot the scaled probability
      density $S^{1-\alpha}~ \rho(\tilde{n}/ S^{1-\alpha}|S)$ versus
      the scaled number of non-zero routes $\tilde{n} / S^{1-\alpha}$,
      combining data from all years.  In agreement with
      (\protect\ref{eq:dist-scaling}), we find that the scaled data
      fall onto a single curve.\label{fig:occup_collapse}}
  \end{center}
\end{figure*}

We can gain further insight into the organization of a species
population in different routes by considering how the distribution of
number of routes with non-zero counts depends on total counts.  That
is, we may quantify the organization of the subpopulations through the
conditional probability density $\rho(\tilde{n}|S)$, which measures
the probability to find a bird species with S total counts having
non-zero counts in $\tilde{n}$ distinct routes.
Fig.~\ref{fig:bbs_units_scaling} suggests that $\rho(\tilde{n}|S)$
will have a peak that increases as a power law with $S$.  As shown in
Fig.~\ref{fig:occup_collapse}a this is indeed the case.  If the data
exhibit scaling, we should be able to identify a universal scaling
function $h$ such that
\begin{equation}
\rho(\tilde{n}|S) \sim \frac{1}{S^{\alpha}} ~~ h
\left(\frac{\tilde{n}}{S^{\alpha}} \right) \,,
\label{eq:dist-scaling}
\end{equation}
We test the scaling hypothesis in Eq.~(\ref{eq:dist-scaling}) by
plotting the scaled variables:
\begin{equation}
S^{\alpha}~ \rho \left( \left . \frac{\tilde{n}}{S^{\alpha}} \right |S
\right) ~~~{vs.} ~~~ \frac{\tilde{n}}{S^{\alpha}} \,. 
\end{equation}
Fig.~\ref{fig:occup_collapse}b shows that all curves collapse onto a
single curve, which yields the scaling function $h(u)$.

\section{Discussion}

Our analysis differs from Taylor's original studies
~\citep{taylor:lr-1961, taylor:lr-woiwod-1982, taylor:lr-1984} in an
important way.  Taylor was interested in comparative analysis and so
calculated a separate exponent for each species.  He did this by
analyzing multiple samples, replicated across time or space, for each
species.  Here, we calculate exponents replicating \emph{across}
species.  One advantage of this approach is that we analyze the time
series of total counts, summed over the entire survey.  These total
counts are considerably more robust estimates of abundance than local
counts taken at individual routes.

Another advantage of analyzing scaling across species is that is
allows us separate general patterns or ``laws'' that are invariant
across taxonomic groups from general rules that may explain
\emph{deviations} from these laws.  Our reasoning is that when the
physical dimensions of a problem, such as energy or material flows, or
spatial population structure, predominate, we should observe scaling
laws that do not depend strongly on the biological differences among
species, but that species-specific differences should appear as
residual variation after the common scaling laws are factored out.
That so many species fall along a single scaling relationship
describing variability as a function of population size
(Fig.~\ref{fig:time_series_scaling}--\ref{fig:tent_shape}) suggests
that there may be universal features to the way in which North
American breeding bird populations are subdivided spatially.  We find
exactly these features in the invariant, 3/4-power scaling of number
of occupied survey routes versus total population size
(Fig.~\ref{fig:bbs_units_scaling}--\ref{fig:occup_collapse}).

However, not all of the variability in the data is accounted for by
these scaling laws.  For example, species with average total counts of
approximately 250 individuals exhibit more than two orders of
magnitude range in growth rate standard deviation
(Fig.~\ref{fig:time_series_scaling}b).  We believe that this residual
variation does reflect important aspects of the ecology of individual
species.  There is a strong correlation between the residuals in
Fig.~\ref{fig:time_series_scaling}b and the area of the corresponding
species ranges, measure in terms of the average number of non-zero
routes (T. Keitt, unpublished results).  A likely explanation for this
pattern is that fluctuations in the abundances of broadly distributed
species will tend to average out spatially because different regions
are influenced by geographically distinct climate regimes.  Thus, it
appears that species whose life histories tend to produce strongly
aggregated distributions (i.e., species that are locally common, but
regionally rare) are the ones that fluctuate the most relative to
their total abundance.  Species that have broad spatial distributions
(i.e., locally rare, but regionally common) are therefore expected to
fluctuate less than similar species with more restricted geographic
ranges.

Ranking species in terms of their residual deviation from the
growth-scale law (Fig.~\ref{fig:time_series_scaling}b) supports our
hypothesis that locally common, but regionally rare species fluctuate
more than expected, and vise versa.  Large positive residuals
correspond to species with restricted geographic ranges, such as
Golden-cheeked Warbler (\textit{Dendroica chrysoparia}; 2.5 times more
variable), species that are habitat specialists and nest in large
colonies, such as Tricolored Blackbird (\textit{Agelaius tricolor};
13.7 times more variable), species that breed in large groups called
``leks'', such as Greater Prairie Chicken (\textit{Tympanuchus cupido};
3 times more variable), and species that show strong local migration
patterns in response to changes in resource availability, such as
White-winged Crossbill (\textit{Loxia leucoptera}, 3.5 times more
variable) and Red Crossbill (\textit{Loxia pytyopsittacus}; 3.7 times
more variable).  Species that show low variability in relation to the
scaling-law are typically solitary, territorial breeders such as
Yellow-throated Warbler (\textit{Dendroica dominica}; 2.8 times less
variable), Prairie Falcon (\textit{Falco mexicanus}; 2.6 times less
variable), Swamp Sparrow (\textit{Melospiza georgiana}; 2.5 times less
variable), Kentucky Warbler (\textit{Oporornis formosus}; 2.4 times
less variable), and Chuck-will's-widow (\textit{Caprimulgus
  carolinensis}; 2.3 times less variable).  The important point is
that had we started from a purely autecological standpoint and
ignored the important physical dimensions of the problem (e.g.,
structure of geographic ranges), we could easily have missed key
pattern in terms of deviations from general scaling laws.

We should however mention several caveats.  We do not as of yet know
whether our results can be generalized to include other, non-avain
taxonomic groups, or to other continents and climate regimes.  Also,
despite our use of highly aggregated, and therefore more robust time
series, we suspect that there remain sources of variation in our
analysis unrelated to actual population fluctuations.  One vexing
problem is repeated local migration between sampled and unsampled
locations (we call this ``sloshing'').  Even if there is no variation
in the true abundance across years, sloshing will lead to a given
individual being counted in some years and not others, leading to
measurement errors in the time series.  We suspect this effect is not
a significant component of variation in most of our time series, but
may be substantial in a few cases.  (Sloshing may contribute to the
extreme variability of Tricolored Blackbird, for example.)  Additional
data, such as mark-recapture, may be need to resolve this issue.

There are two additional mechanisms related to our model for
geographically subdivided populations that we have not discussed.  We
have shown how a non-linear dependency of the number of subpopulations
versus total population size may explain the observed deviation from
1/2-power scaling of population fluctuations.  Our basic hypothesis
depends on the average local abundance scaling with total abundance in
independently fluctuating subpopulations of roughly equal size.
However, there are other patterns that may influence the ``effective''
number of independently fluctuating subpopulations and thus partially
amount for the observed exponent in the growth-scaling law.  First,
large spatial variation in local abundance~\citep{brown-etal-1995}
could cause wide-spread species to fluctuate with greater magnitude
than if all subpopulations have the same local abundance, since most
of the variation would be driven by a few, high abundance sites.
Second, strong spatial autocorrelation in population growth increments
or ``spatial synchrony'' among fluctuating
subpopulations~\citep{grenfell-etal-1998, bjornstad-etal-1999,
  kendall-etal-2000, lundberg-etal-2000} may also cause a reduction in
the effective number of independent subpopulations, and thus account
for the increased magnitude of fluctuation in broadly distributed
species.  Temporal autocorrelation may act similarly to increase or
decrease variability relative to our model.  The consequences of these
mechanisms need further exploration.

A surprising result of our analysis is the, to our knowledge,
previously unreported 3/4-power scaling of spatial distribution as a
function of total population size (Fig.~\ref{fig:bbs_units_scaling}).
This result is closely related to, but not the same as, the
``Distribution-Abundance'' curve of \cite{hanski-gyllenberg-1997} that
describes the fraction of regional habitats occupied as a function of
average local abundance.  We do not as of yet have an explanation for
why the exponent should take this particular value, nor why it is so
consistent through years.  Recently, there has been considerable
interest in explanations for the apparent 3/4-power scaling law
relating body mass to metabolic output~\citep{enquist-etal-1998,
  west-etal-1999, dodds-etal-2001, niklas-enquist-2001}.  One
explanation posited to explain 3/4-power scaling is optimal
structuring of a fractal transport network, such as the vascular
system of plants and animals~\citep{west-etal-1999}.  This suggests an
interesting hypothesis to explain 3/4-power scaling in our analysis:
if the geographic ranges of species are subdivided in according to a
particular fractal pattern, perhaps because of the fractal nature of
the physical environment~\cite[e.g.,][]{rinaldo-etal-1995}, then it
might lead to our observed scaling laws.  Testing this hypothesis will
require additional study.

It is interesting to note that our results are in striking qualitative
agreement with similar studies from a broad range of social systems,
ranging from growth of companies in the U.S. economy to the GDP of
countries~\citep{stanley:mhr-etal-1996, lee-etal-1998,
  plerou-etal-1999a}, suggesting that our simple model of growth may
apply quite broadly~\citep{amaral-etal-1998}.  Our observation that
more ``specialized'' birds (in terms of smaller number of
subpopulations) fluctuate more in total number than those that average
fluctuations over many subpopulations may have an interesting parallel
in social organizations: those that specialize on a few economic
activities, e.g., countries with a single export product, may
fluctuate considerably more than similarly sized organization with
diverse economic activities, e.g., countries that produce a range of
products.  Putting all of one's eggs in a single basket, as the saying
goes, some times leads to catastrophes, and, it appears, greater
variability as well.

\section{Acknowledgements}

T. Keitt thanks the Santa Fe Institute and the National Center for
Ecological Analysis and Synthesis for support during the initial phase
of this research.  This research was made possible by the efforts by
thousands of U.S. and Canadian BBS participants in the field, as well
as, USGS and CWS researchers and managers.

\section*{Appendix: Corrections applied to time series}

Let $s_{tuv}$ be the number of birds of species $u$ counted at route
$v$ in year $t$.  The raw total counts $R_{tu} = \sum_v^{N_t} s_{tuv}$
contain information about the abundance of species $u$ in year $t$ as
well as information about the number $N_t$ and distribution of routes
surveyed in year $t$.  The goal is to remove the bias in the counts
$R_{tu}$ introduced by variation in the number and distribution of
survey routes through time.  We do this by replacing each count
$s_{tuv}$ for a given species at a given route with the time average
$\mu_{uv} = T_v^{-1}\sum_ts_{tuv}$ for that route and species, where
$T_v$ is the number of years that route $v$ was surveyed.  We then
construct new, surrogate time series $M_{tu} = \sum_v^{N_t} \mu_{uv}$
whose variation only reflects changes in the number and distribution
of survey routes through time (because the same $\mu_{uv}$ is used in
each year), and not any real change abundance.  We can then generate a
bias corrected time series by subtracting these new time series from
the raw totals:
\begin{equation}
  S_{tu} = R_{tu} - M_{tu} + \bar{M}_u
\end{equation}
where $\bar{M}_u$ is the time average of $M_{tu}$ for species $u$.
The advantage of this approach is that survey routes added or removed
outside a species range will not influence the corrected total,
because these routes will have $\mu_{uv} = 0$.

\bibliography{research}
\bibliographystyle{natbib}

\end{document}